# "Topological" Hall effect in the ferromagnetic microparticles with noncoplanar magnetization distribution

O.G. Udalov, A.A. Fraerman

Institute for physics of microstructures RAS, 603950, Russia, Nizhny Novgorod, GSP-105

It is well known that there is a contribution in ferromagnetics to the transverse conductivity linear in magnetization, so-called abnormal Hall effect (AHE). AHE occurs due to spin-orbit interaction of current carriers. At the present time it is demonstrated that in the ferromagnetics with inhomogeneous magnetization an additional contribution to transverse conductivity can occur. At that the nature of the effect is the exchange interaction between electrons in the media. Sometime this effect is called as "topological" Hall effect (THE). THE is observed experimentally in $Nd_2M_2dO_7$ [1] and also in MnSi [2]. Microscopic theory of the THE in the crystals of magnetic insulator is given in Ref. [3]. Simple semiclassical explanation of the THE is provided by the paper of Aharonov and Stern [4]. In Ref. [5] it is shown theoretically that the exchange interaction can leads also to persistent currents in the mesoscopic ring as in the Aharonov-Bohm effect, if magnetization distribution is nonuniform and noncoplanar. Recently Bruno [6] proposed the artificial system for THE observing. The system was the magnetic semiconductor placed in the nonuniform magnetic field of ferromagnetic nanodipoles array. In the present paper we consider another artificial system, in which THE should exist, namely ferromagnetic nanoparticles with "vortex" and "antivortex" distribution of magnetization.

Phenomenological theory for conductivity of nonuniform magnetic media is proposed. The theory is based on some assumptions concerning investigated media. The main of them is that the exchange interaction dominates in the media and all relativistic interactions can be neglected. In this case one says about exchange symmetry of the system. Exchange interaction depends only on the mutual orientation of spins, therefore simultaneous rotation of all the spins in the system on the same angle do not change Hamiltonian. Correspondingly such a parameter as conductivity also should not depend on the all spins rotation (magnetization rotation). Another assumption is that the magnetization in the medium has the same magnitude in all the point. At that variation of magnetization direction in the media is of long range comparing to lattice period and electron free path length. In this case conductivity can be presented as a sum of uniform media conductivity and small correction caused by magnetization heterogeneity. So, for the current in the media placed in the electric field $\vec{E}$ one has:

$$\vec{j} = \sigma_0(\vec{E}, |\vec{M}|)\vec{E} + \sigma_M(\vec{M}, \frac{\partial}{\partial \vec{r}}\vec{M}, ...)\vec{E}. \qquad (1)$$

Obviously, tensor $\sigma_M$ should contain even number of spatial derivatives. The simplest tensor meeting the requirements mentioned above has the form:

$$\sigma_{i,j}^{MR} = \hat{\alpha}_{ijkl}^1 \left( \frac{\partial \vec{M}}{\partial x_k} \cdot \frac{\partial \vec{M}}{\partial x_l} \right), \qquad (2)$$

where $\hat{\alpha}_{ijkl}^1$ has a symmetry of lattice. Tensor (2) describes well known giant magnetoresistance effect.

Only the tensor including three magnetization vectors, two spatial derivatives and is not changed under magnetization rotation is determined by the expression:

$$\sigma_{i,j}^H = \hat{\alpha}_{ijkl}^2 \left( \vec{M} \cdot \left[ \frac{\partial \vec{M}}{\partial x_k} \times \frac{\partial \vec{M}}{\partial x_l} \right] \right). \qquad (3)$$

For the isotropic media tensor (3) can be rewritten as:

$$\sigma_{i,j}^H = K_H \left( \vec{M} \cdot \left[ \frac{\partial \vec{M}}{\partial x_i} \times \frac{\partial \vec{M}}{\partial x_j} \right] \right). \qquad (4)$$

The $\sigma^H$ is an antisymmetric tensor because of vector product properties. Thus it describes the effect of Hall kind. Natural to assume that (4) describes "topological" Hall effect. It is seen from (4) that THE is non zero only in the medium with noncoplanar magnetization distribution. Moreover magnetization distribution should vary along two directions.

In the paper we use obtained phenomenological expression for analysis THE in the particles with "vortex" and "antivortex" magnetization distribution.

"Vortex" and "antivortex" magnetization distribution are described by the formula:
$$\vec{M} = (\pm\sqrt{1 - m_z^2(\rho)} \sin(\varphi), \sqrt{1 - m_z^2(\rho)} \cos(\varphi), m_z(\rho)), \qquad (5)$$
where $\rho, \varphi, z$ are cylindrical coordinates. Sign "−" corresponds to "vortex" state, and "+" corresponds to "antivortex". Substituting (5) into (4) one can be sure that tensor $\sigma_{i,j}^H$ has only two nonzero components:
$$\sigma_{x,y}^{HV} = -\sigma_{y,x}^{HV} = \pm K_H \frac{1}{\rho} \frac{\partial m_z(\rho)}{\partial \rho}. \qquad (6)$$

At that two mentioned magnetization distributions give different signs in front of $K_H$. The "vortex" state is realized in the ferromagnetic particles of cylindrical shape [7]. Thickness of the particles is about 30 nm and diameter is in the range from 50 to 500 nm. The realization of "antivortex" state in magnetic particles is more complicated problem, therefore further only cylindrical particles with "vortex" magnetization will be considered.

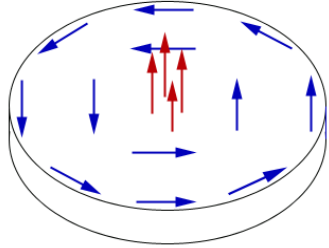

Fig. 1. The "vortex" state in the magnetic nanodisk.

Averaging (6) over cylindrical ferromagnetic particle with "vortex" gives the following expression for THE:
$$\overline{\sigma_{x,y}^{HV}} = -\overline{\sigma_{y,x}^{HV}} = \int dx dy \sigma_{x,y}^{HV}(x,y) = K_H (M_z(R_d) - M_z(0)). \qquad (7)$$

Here $R_d$ is particle radius. As it is seen from (7) THE depends only on the difference between z-component of magnetization in the center and at the edge of the particle. At the absence of external magnetic field magnetization at the particle edge lies in the disk plane. Therefore $M_z(R_d) = 0$. In the center of the particle (in the region of vortex core) magnetization is perpendicular to the disk plane (see fig. 1). When external magnetic field applied along the z axis and co-directed with core magnetization, later is not changed, but at the edge of the particle z-component of magnetization is increased. Therefore THE goes down with magnetic field increasing. If external magnetic field direction is opposite, THE increases until core magnetization change it's sign. Qualitative dependence of THE in the cylindrical ferromagnetic particle with vortex magnetization on external magnetic field applied along z-axis is presented in Fig. 2.

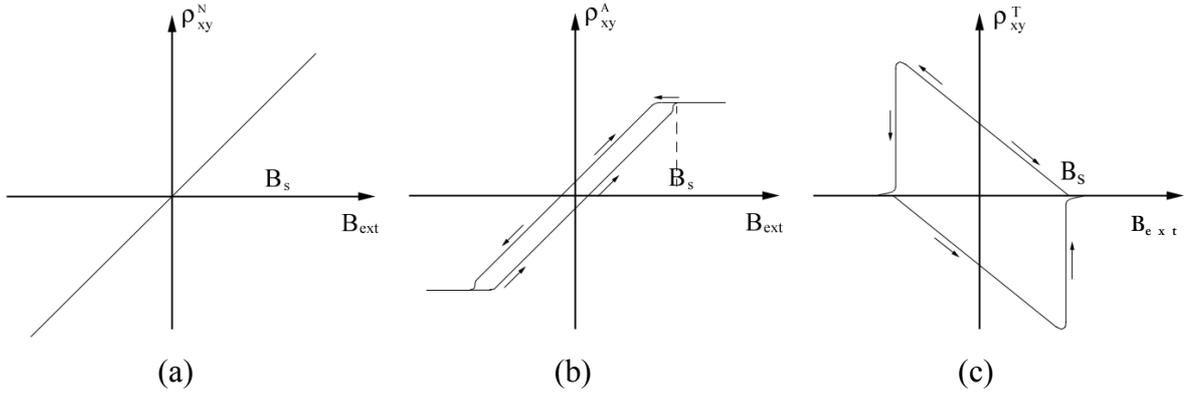

Fig. 2. Hysteresis loop for normal (a), abnormal (b) and "topological" Hall (c) effects.

As it is seen from Fig. 2 hysteresis loop for THE is qualitatively different for normal and abnormal Hall effects hysteresis loops.

Using reasoning of the previous paragraph it is easy to understand the dependence of the THE on the magnetic field applied in the plane of the particle (Fig. 3).

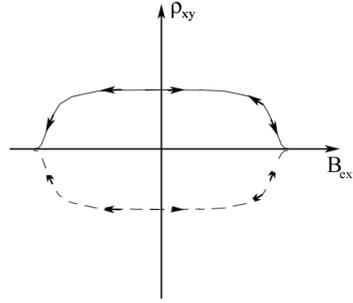

Fig. 3. Dependence of the THE on the magnetic field applied in the plane of the disk.

The estimation of the THE magnitude in ferromagnetic nanodisk can be done using of simple semiclassic model. Let us propose that s-d model describes transport properties of investigating particle. Then conduction electron description can be done in the frame of semi classic model proposed by Aharonov and Stern [4]. Using formula (11) from Ref. [4] the effective "topological" magnetic field acting on the electron can be estimated. For the "vortex" particle "topological" magnetic field is directed along z-direction. It's magnitude is estimated as:

$$B_{eff} \ll \frac{\Phi_0}{L^2}.$$

Here $\Phi_0$ is magnetic flux quantum, $L$ is a diameter of the particle. Note that "topological" field has an opposite sign for opposite electron spin orientations. Therefore THE has an opposite sign for two conduction electron spin subbands and it should be proportional to the ratio of spin subbands splitting and Fermi energy ($J/\varepsilon_f$). For Co particle with diameter of 100 nm $B_{eff} \approx \frac{\Phi_0}{L^2} \approx 2000\ Oe$ and $\frac{J}{\varepsilon_f} = 0.1$. Correspondingly for THE resistance $\rho_X^{HV}$ one has:

$$\rho_X^{HV} = R_0 B_{eff} \approx \frac{\Phi_0}{L^2} \frac{J}{\varepsilon_f} R_0 = 1.6 \cdot 10^{-10}\ \text{Ohm·cm}.$$

Here $R_0$ is the constant of normal Hall effect.

Note that normal and abnormal Hall effects also contribute to the transverse resistivity. The abnormal Hall effect is linear in magnetization. In the absence of external magnetic field the average magnetization of "vortex" particle is directed along z-axis and is determined only by the region of core. The zero field abnormal Hall effect resistivity can be estimated by the formula:

$$\rho_X^A = R_S M_S \frac{L_c^2}{L^2} = 0{,}8 \cdot 10^{-11} \text{ Ohm·cm,} \quad (3.12)$$

where $M_S$ is the saturation magnetization of the particle material, $L_c$ is "vortex" core lateral size (about 20 nm).

At the presence of external magnetic field $B_{ext}$ is directed along the z-axis the normal Hall effect contributes to the transversal resistivity.

$$\rho_{xy}^N = R_0 B_{ext}. \quad (3.13)$$

Thus THE gives dominant contribution to the transverse resistivity if the external field less than 200 Oe ($B_{eff} J / \varepsilon_F > B_{ext}$). The hysteresis loop for Co nanodisk (magnetic field is perpendicular to the disk plane) is presented on the Fig. 4.

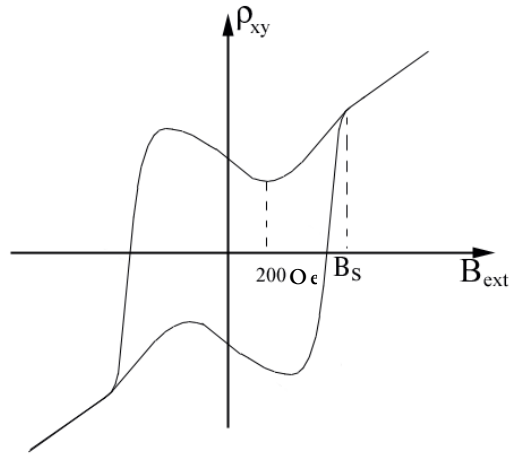

Fig. 4. Hysteresis loop for Co nanodisk (100 nm diameter). $B_S$ is saturation field.

Thus in this paper the phenomenological theory for "topological" Hall effect is proposed. The theory can be used for qualitative analysis of THE in the medium with arbitrary magnetization distribution. In the paper the cylindrical particle with "vortex" magnetization is considered. The hysteresis loop of THE is qualitatively different comparing to abnormal Hall effect. The estimation of THE magnitude in the "vortex" particle is made. For Co particle with 100 nm diameter the THE exceeds normal and abnormal Hall effects until an external magnetic field less than 200 Oe.

References

[1] Y. Taguchi, Y. Oohara, H. Yoshizawa, et. al, Science, 291, 2573 (2001).
[2] A. Neubauer, C. Pfleiderer, B. Binz, et. al, Phys. Rev. Lett., 102, 186602 (2009)
[3] Shigeki Onoda and Naoto Nagaosa, Phys. Rev. Lett., 90, 19, 196602-1 (2003).
[4] Ya. Aharonov, A. Stern, Physical Review Letters, 69, 25, 3593 (1992).
[5] G. Tatara, H. Kohno, Phys. Rev. B , 67, 113316 (2003).
[6] P. Bruno, V. K. Dugaev, and M. Taillefumier, Phys. Rev. Lett., 93, 096806-1 (2004).
[7] R.P. Cowburn, D.K. Koltsov, A.O. Adeyeye et al., Phys. Rev. Lett., 83, 1042 (1999)